%% file: piep.tex
\documentclass[12pt,twoside,a4paper,leqno]{article}

\usepackage[latin1]{inputenc}
\usepackage[T1]{fontenc}
\usepackage{parskip}

\usepackage{amsmath,amsfonts,amssymb,amsxtra}
\usepackage{latexsym}
\usepackage{theorem}
\usepackage{graphicx,psfrag}
\usepackage{verbatim}
\usepackage{longtable}
\usepackage{fancyhdr}

\author{T.~Dudas}
\title{Spacetime Warps for Spinning Particles Possible?}

{\theoremstyle{break}             
\setlength{\theorempreskipamount}{1cm}
\setlength{\theorempostskipamount}{1cm}
\theorembodyfont{\rmfamily}
\newtheorem{Definition}{Definition}

{\theoremstyle{break}                       
\theorembodyfont{\rmfamily}
\newtheorem{Theorem}[Definition]{Theorem}}

{\theoremstyle{break}                       
\theorembodyfont{\rmfamily}
\newtheorem{Lemma}[Definition]{Lemma}}

{\theoremstyle{break}                       
\theorembodyfont{\rmfamily}
}

{\theoremstyle{break}                       
}

{\theoremstyle{break}                       
\theorembodyfont{\rmfamily}
}

\begin{document}    

\maketitle
\begin{abstract}
By incorporating spinning particles into the framework of\\
 classical General Relativity,
the theory is changed insofar, as, though using holonome coordinates, the connexion becomes asymmetrical.\\
This implies that partial derivatives do not commute any longer, necessarily.\\
Hence, the class of functions under consideration has to be extended.\\
A non minimal extension leads to the possibility of spacetime warps for spinning particles.
\end{abstract}

\input{piep1}

\end{document}

%% file: piep1.tex
The asymmetry of a spinning particle´s energy-momentum tensor is well-known from Noether´s theorem [2].
\begin{Theorem}
The asymmetry of the energy-momentum tensor of the spin particle leads to the asymmetry of the connexion.\\
\begin{equation}
T^{\mu \nu} \not= T^{\nu \mu} \Longrightarrow \; \Gamma^ {\mu \nu} \not=  \Gamma^ {\nu \mu}
\end{equation}
\end{Theorem}

\subsubsection*{Proof}

Let us subdivide the proof into differrent lemmata.\\

First, we shall show that the asymmetry of the energy-momentum tensor yields to to the asymmetry of the Einstein tensor.\\
This leads to the asymmetry of the Ricci tensor and that yields to the asymmetry of the connexion.\\
\begin{equation}
T_{\mu \nu} \not= T_{\nu \mu} \,\overset{\!\!\!(1)}\Longrightarrow \, G_{\mu
\nu} \not= G_{\nu \mu} \,\overset{\!\!\!(2)} \Longrightarrow \, R_{\mu \nu}
\not= R_{\nu \mu} \,\,\overset{\!\!\!(3)} \Longrightarrow
\Gamma^\lambda{}_{\mu \nu} \not= \Gamma^\lambda{}_{\nu \mu}.
\end{equation}
The third part of the proof we shall show indirectly.
\begin{equation}
\Gamma^\lambda{}_{\mu \nu} = \Gamma^\lambda{}_{\nu \mu}\,\,\overset{\!\!\!(3)} \Longrightarrow
 R_{\mu \nu} = R_{\nu \mu}
\end{equation}
\begin{Lemma}
\begin{equation}
T_{\mu \nu} \not= T_{\nu \mu} \,\overset{\!\!\!(1)}\Longrightarrow \, G_{\mu
\nu} \not= G_{\nu \mu}
\end{equation}
\end{Lemma}

\subsubsection*{Proof}
The asymmetry of the energy-momentum tensor yields to to the asymmetry of the Einstein tensor. This is shown most directly by the field equations.
\begin{equation}
G_{\mu \nu} = \frac{1}{8\pi}\, T_{\mu \nu}
\end{equation}

\begin{Lemma}
\begin{equation}
 G_{\mu \nu} \not= G_{\nu \mu} \,\overset{\!\!\!(2)} \Longrightarrow \, R_{\mu
\nu} \not= R_{\nu \mu}
\end{equation}
\end{Lemma}

\subsubsection*{Proof}
If the Einstein tensor is asymmetrical, either the Ricci tensor has to be asymmetrical or the metrical tensor, as can be
concluded by the definition of the Einstein tensor.
\begin{equation}
G_{\mu \nu} = R_{\mu \nu} - \frac{1}{2} \,g_{\mu \nu} \,R
\end{equation}

As the metrical tensor is not asymmetrical due to its transformation\\
property, the Ricci tensor remains the only candidate.\\
The metrical tensor transforms as follows:\\
\begin{equation}
g_{\mu \nu} = \frac{\partial \xi^\alpha}{\partial\, x^\mu} \, \frac{\partial
\xi^\beta}{\partial\, x^\nu} \: \eta_{\alpha \beta} ,
\end{equation}
due to the invariance of the infinitesimal line element $ds^2= g^{\mu \nu}dx^\mu dx^\nu$.\\
$\eta_{\alpha \beta}$ designates the metrical tensor of flat space.\\

The metrical tensor cannot be asymmetrical, as real numbers commute and if the matrix elements
of the metrical tensor are not forced by hand to be matrices of higher rank, they are real
numbers.\\
As $g_{\mu \nu}$ is not asymmetrical, $R_{\mu \nu}$ has to be.\\
\begin{Lemma}
\begin{equation}
 R_{\mu \nu} \not= R_{\nu \mu} \,\,\overset{\!\!\!(3)} \Longrightarrow
\Gamma^\lambda{}_{\mu \nu} \not= \Gamma^\lambda{}_{\nu \mu}.
\end{equation}
\end{Lemma}

\subsubsection*{Proof}
This proof is indirect, as well, which means we show that the symmetry of the connexion leads to the symmetry of the Ricci tensor.\\
\begin{equation}
\Gamma^\lambda{}_{\mu \nu} = \Gamma^\lambda{}_{\nu \mu}\,\,\, \Rightarrow \,
R_{\mu \nu} = R_{\nu \mu} .
\end{equation}
As this part of a part of the proof is not straightforward, we shall subdivide
it into different parts.\\
\begin{eqnarray}
\hspace*{2cm}\Gamma \,^ \lambda{}_ {\mu \nu} = \,\Gamma\,^\lambda {}_{\nu
\mu}\; & \overset{\!\!\!(3.1)}\Longrightarrow & \,\, \Gamma \,^\lambda {}_ {
\mu \nu} = \frac {1}{2} \,g \,^{\lambda \varrho} \,(\partial _ \mu\, g_{\nu
\varrho} + \partial _ \nu\, g_{\varrho \mu} - \partial \,_ \varrho g_{\mu
\nu})\\
                                                                    &
\overset{\!\!\!(3.2)}\Longrightarrow &  \,\, R_{ \mu \nu} = R_{\nu \mu}
\end{eqnarray}
The symmetry of the connexion leads to the well-known correlation between the Christoffelsymbol and
the metrical tensor.\\
But if the connexion has the form of the Christoffelsymbol, the Ricci tensor is symmetrical.
\subsubsection*{Proofs of 3.1 and 3.2}
The proofs can be reconsidered in [1].\\

Therefore, we have shown that the symmetry of the connexion leads to the symmetry of the Ricci tensor and
hence, that the asymmetry of the Ricci tensor yields to the asymmetry of the connexion by logical
denial.$\Box$\\

\begin{Theorem}
If the connexion is asymmetrical in holonome coordinates, the partial derivatives do not commute any longer.\\
\begin{equation}
\Gamma^ {\mu \nu} \not=  \Gamma^ {\nu \mu}  \;\Longrightarrow \;
\frac{\partial^2 x^\lambda}{\partial x_\mu \partial x_\nu} \not=  \frac{\partial^2 x^\lambda}{\partial x_\nu \partial x_\mu}
\end{equation}
\end{Theorem}

\subsubsection*{Proof}
The equation of motion of a scalar particle in holonome coordinates is[1]
\begin{equation}
  \frac{d^2x^{\lambda}}{d\tau^2} +
                              \frac{\partial x^\lambda}{\partial \xi^\alpha}
                              \frac{\partial^2 \xi^\alpha}{\partial x^\nu \partial x^\mu}
                            \; \frac{dx^\mu}{d\tau}\frac{dx^\nu}{d\tau}
                 = 0
\end{equation}
This yields to to the following equation
\begin{Definition}
\begin{equation}
\frac{\partial x^{\lambda}}{\partial \xi^{\alpha}}
                              \frac{\partial^2 \xi^{\alpha}}{\partial x^{\nu} \partial x^{\mu}} = \Gamma^\lambda{}_{\nu \mu}
\label{connexion}
\end{equation}
\end{Definition}
Therefore, if the connexion is asymmetrical, partial derivatives do not commute any longer.\\

\begin{Theorem}
As a consequence, the class of functions under consideration has to be extended (Analysis2).\\
\begin{equation}
x^\mu \epsilon \; X \subset Y
\end{equation}
\end{Theorem}

\subsubsection*{Proof}
This is clear: The partial derivatives commute for continuous functions, the derivatives of which are continuous[3].\\
Therefore, if the partial derivatives do not commute any more, the functions under consideration are either
not continuous or their derivatives are not.\\
We might as well assume, their environment of definition is not open, which means that Cauchy-chains
do no longer converge or that they do not exist at all.\\
In any case, the class of functions under consideration has to be extended.\\

\begin{Theorem}
This makes non-minimal extensions possible, which give rise to spacetime warps.\\
\begin{equation}
 Y \subset\; Z,\; z\; \epsilon \;Z  \;\Longrightarrow \; z \;\textnormal{not continuous}
\end{equation}
\end{Theorem}

\subsubsection*{Proof}
We have to extend the class of functions and the minimal extension would be to assume,
that derivatives of the function are no longer continuous.\\
But this is not the only possibility.\\
We might as well assume, the function itself is not continuous and in the case of \;(~\ref{connexion}),
this indeed implies the existence of spacetime warps for spinning particles.\\

\newpage
\section*{Literature}
[1] S.Weinberg: Gravitation and Cosmology. Wiley \& Sons 1972

[2] J.D.Bjorken, S.D.Drell: Relativistic Quantum Fields. McGraw-Hill 1965

[3] O.Forster: Analysis2. Vieweg, Braunschweig 1981